\documentclass[twocolumn,showpacs,prb]{revtex4}

\usepackage[dvips]{graphicx}
\usepackage{amsmath,amssymb}

\newcommand{\chem}[1]    {${\mathrm{#1}}$}

\newcommand{\CNCMO}{\chem{Nd_{0.5}Ca_{0.5}Mn_{1-\mathit{z}}Cr_\mathit{z}O_3\ (\mathit{z}=0.03)}}
\newcommand{\NCMO}{\chem{Nd_{0.65}Ca_{0.35}MnO_3}}
\newcommand{\PCMO}{\chem{Pr_{0.7}Ca_{0.3}MnO_3}}
\newcommand{\NSSMO}{\chem{(Nd_{1-\mathit{y}}Sm_\mathit{y})_{0.5}Sr_{0.5}MnO_3\ (\mathit{y}=0.75)}}
\newcommand{\LPCMO}{\chem{(La_{1-\mathit{y}}Pr_\mathit{y})_{0.7}Ca_{0.3}MnO_3\ (\mathit{y}=0.7)}}

\def\btt#1{\texttt{\@backslashchar#1}}%
\DeclareRobustCommand\bblash{\btt{\@backslashchar}}
\makeatother

\begin{document}

\preprint{HEP/123-qed}

\title[Short Title]{Multivalued memory effects in electronic phase-change manganites controlled by Joule heating}%

\author{H. Song}
\email{song@pothos.t.u-tokyo.ac.jp}
\author{M. Tokunaga}%
\author{S. Imamori}%
\author{Y. Tokunaga}%
\author{T. Tamegai}%
\affiliation{%
Department of Applied Physics, The University of Tokyo, 7-3-1 Hongo, Bunkyo-ku, Tokyo 113-8656, Japan }%

\date{\today}
\begin{abstract}
Non-volatile multivalued memory effects caused by magnetic fields, currents, and voltage pulses are studied in \NCMO\ and \NSSMO\ single crystals in the hysteretic region between ferromagnetic metallic and charge-ordered insulating states. The current/voltage effects observed in this study are explained by the self-heating effect, which enable us to control the colossal electroresistance effects. This thermal-cycle induced switching between 
electronic solid and liquid states can be regarded as electronic version of atomic crystal/amorphous transitions in phase-change chalcogenides. 
\end{abstract}

\maketitle
Competition between different electronic phases plays a crucial role in extracting extraordinary changes by external perturbations~\cite{Imada}. In manganites, competing ferromagnetic metal (FMM) and charge-ordered insulator (COI) phases provide a playground for the colossal magnetoresistance effects~\cite{Tokurabook}. Due to the interaction between spin, charge, lattice, and orbital degrees of freedom, drastic changes in resistance can also be induced by application of pressure~\cite{Moritomo}, electric fields~\cite{Asamitsu}, X-ray~\cite{Kiryukhin}, or visible light~\cite{Miyano}. The sensitivity to these external perturbations increases as the system approaches the phase boundary. Near the phase boundary, introduction of randomness, which is inevitable in solid solutions like this class of manganites, causes instability to the phase separation~\cite{Dagotto}.

One of the canonical phase-separated (PS) systems is achieved by Cr-substitution for Mn in charge-ordered manganites. Kimura $et~al.$ reported that field-induced changes in Nd$_{0.5}$Ca$_{0.5}$Mn$_{1-z}$Cr$_z$O$_3$ are frozen even after removing the magnetic field~\cite{Kimura}. This kind of persistent memory effects are currently attracting much interest due to potential applications as non-volatile memory devices~\cite{Liu}. The persistent switching behavior in resistance related with transitions between FMM and COI states has been demonstrated by applying magnetic fields~\cite{Levy}, gate voltages~\cite{Bhattacharya}, or visible light~\cite{Takubo}.

In this paper, we show another route to control the phase fraction of the PS state, that is, Joule self-heating. In our previous study, we demonstrated a collapse of percolative conduction paths in PS states by  the application of a large amount of current, and explained this phenomenon based on the self-heating effect~\cite{Tokunaga1}. Recently, Mercone $et~al.$ revealed that such a heating effect is almost inevitable especially when the resistivity has a negative temperature coefficient~\cite{Mercone}. They also raised  questions to many experimental reports about electric field or current effects on manganites, and claimed that Joule heating effects were underestimated in general. The essential role of the heating effect on transport properties in manganites has been recognized only recently~\cite{Tokunaga1,Mercone,Wu}. Since the heating effect is easy to control, we can finely tune the PS state to derive colossal magnetoresistance effects at low magnetic fields and the current oscillation at constant voltages~\cite{Tokunaga2}. In the present study, we demonstrate a non-volatile memory effect based on the self-heating effect. Since the change in response is large, different values of resistance can be stably achieved. In addition, we show that this memory effect is not only observed in canonical phase-separated manganites such as Cr-doped systems, but also in various manganites showing hysteretic behavior in the phase diagram.

\begin{figure}[!tb]
\begin{center}
\includegraphics[width=7.5cm]{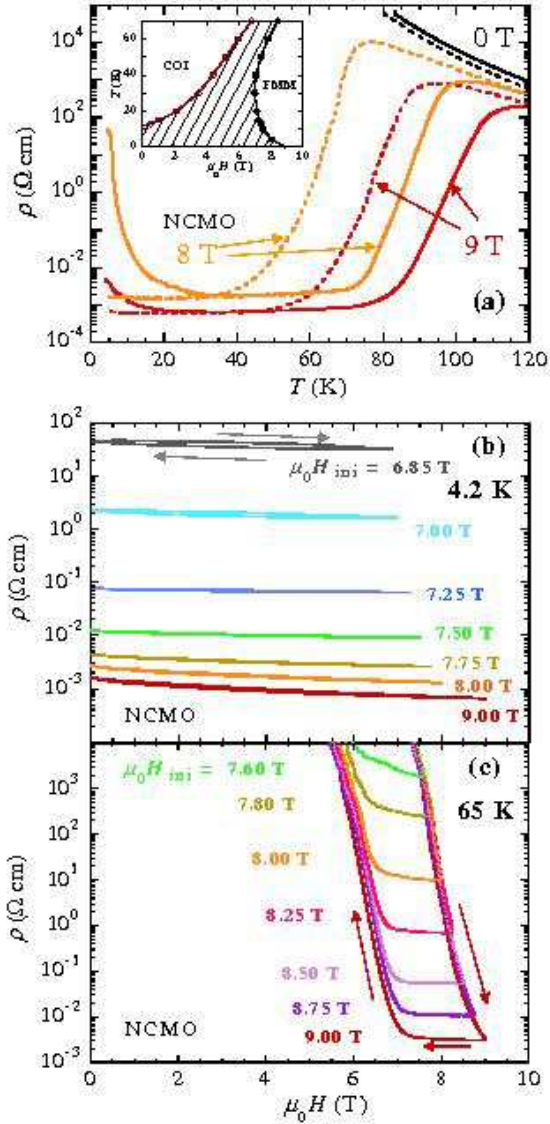}
\caption{(a) Temperature dependence of resistivity in the NCMO crystal at several fields. The solid lines indicate resistivity measured in the warming process after zero-field cooling, while the dashed lines correspond to those obtained in the successive cooling process in magnetic fields. Inset shows the phase diagram in the temperature-field plane. Magnetic field dependence of resistivity at (b)~4.2~K and (c)~65~K was measured after cooling in various fields of $-H_\mathrm{ini}$.}
\label{fig1}
\end{center}
\end{figure}

Crystals of \NCMO~(NCMO) and \NSSMO~(NSSMO) were grown by the floating-zone method. The resistivity ($\rho$) was measured by the standard four-probe method. Four electrodes on the sample were formed by heat-treatment-type silver paste. Detailed  description of the experimental configuration is presented in our previous publications~\cite{Tokunaga1,Tokunaga2}.
Figure~1(a) shows temperature dependence of resistivity in NCMO at fields $\mu _0 H = 0$, 8, and 9~T. Solid lines represent the results obtained in the warming processes after zero-field cooling (ZFC), while the dashed lines show those obtained in the successive field-cooling (FC) processes. At 8~T after ZFC, the $\rho $ first decreases upon warming, and then increases again above 70~K. On the other hand, in the FC process, the $\rho $ decreases
monotonically from 75~K. As a result, the $\rho $ at 5~K after ZFC is nearly five orders of magnitude as large as that after FC. According to an earlier report \cite{Liu2}, NCMO shows ferromagnetic components even at zero field. In addition, as seen in the $\rho-T$ curve at 9~T after ZFC [Fig.~1(a)], magnetic field of 9~T is insufficient to drive the whole sample into FMM state, making it difficult to define the phase boundary between COI and FMM. Here we chose a threshold resistivity of $\rho = 3~\Omega\mathrm{cm}$ to separate $insulating$ and $metallic$ states, and constructed a phase diagram for this crystal from temperature and field dependence of $\rho $ [inset of Fig.~1(a)]. In this figure, closed (open) circles represent the boundary field and temperature from insulator (metal) to metal (insulator). Hatched area corresponds to the hysteretic region. 
Such a broad hysteresis region at low temperatures is characteristic of 
the first-order phase transitions in manganites~\cite{Kuwahara}.

The memory effect clearly shows up in magnetoresistance after field cooling. Figures~1(b) and (c) show magnetic field dependence of resistivity at 4.2~K and 65~K, respectively. Prior to each measurement, the crystal was set in an initial state by cooling in a certain magnetic field ($-H_\mathrm{ini}$). Then we measured magnetoresistance for $H = 0 \to +H_\mathrm{ini} \to 0$. At 4.2~K, each $\rho-H$ curve is well separated: $\rho$ is merely determined by $H_\mathrm{ini}$ as long as $|H|<|H_\mathrm{ini}|$. This memory effect originates from the irreversibility in the phase transition: field induced FMM components cannot be reverted to COI state by reducing the field at this temperature. On the other hand, at 65~K, since COI state is stable at low fields as seen in the inset of Fig.~1(a), every $\rho-H$ curve coincides to each other at low fields. Namely, the memory is reset by removing the field. The observed memory effects at low temperatures are similar to those reported in a canonical PS manganite of Cr-doped \chem{Nd_{0.5}Ca_{0.5}MnO_3} \cite{Kimura} and also in some other manganites~\cite{Gordon,YTokunaga}.

\begin{figure}[!tb]
\includegraphics[width=7.5cm]{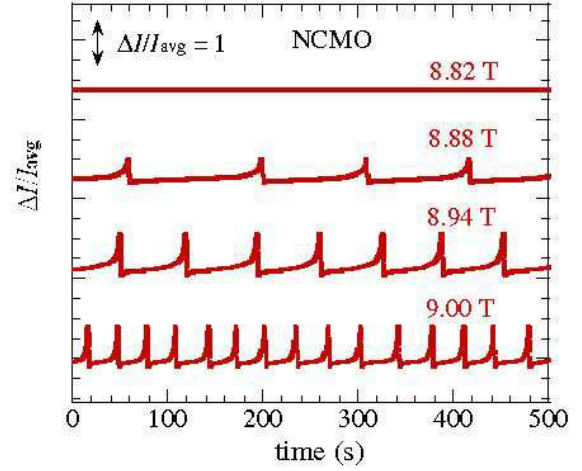}
\caption{Temporal variation of currents in the NCMO crystal at 14~V and 65~K under various magnetic fields. $I_\mathrm{avg}$ and $\Delta I$ denote the average value of current and the deviation from $I_\mathrm{avg}$, respectively.}\label{fig2}
\end{figure}  

The high-field state of NCMO shows an anomalous transport property which is similar to that of canonical PS manganites \LPCMO\ and \CNCMO: current oscillation at a constant voltage~\cite{Tokunaga2}. In our previous study, we 
concluded that the periodic oscillation is characteristic of the first-order transition, related with the closing and opening of conduction paths in inhomogeneous manganites~\cite{Tokunaga2}. In this study, we observe magnetic-field induced percolative FMM states and current oscillations in NCMO.
Figure~2 shows temporal variations of currents at $V = 14$~V and 65~K in various magnetic fields. With increasing the magnetic field, a steep increase in current appears at a certain interval. This result suggests that we can prepare the PS state by application of magnetic field in NCMO. Such a PS scenario in magnetic fields is consistent with early neutron diffraction measurements in \PCMO~\cite{Radaelli,Yamada}. 

\begin{figure}[!tb]
 \includegraphics[width=7.5cm]{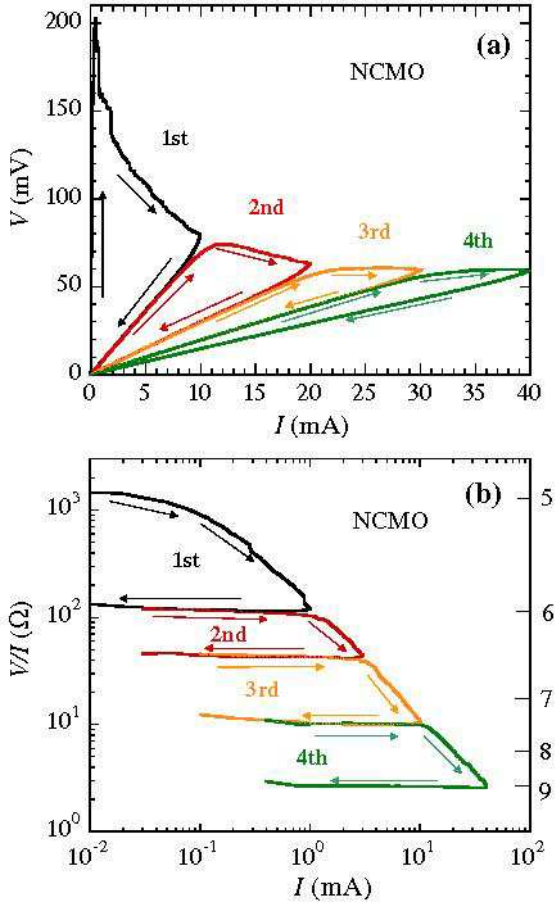}
\caption{(a) $V-I$ curves of the NCMO crystal at 4.2~K and 8~T after ZFC. Each curve is taken with successively increasing $I_\mathrm{max}$ from 10~mA to 40~mA. (b) Current dependence of the $V/I$ and the effective sample temperature ($T^*$) calculated from temperature dependence of resistance measured at $I = 10\ {\rm \mu}$A. The curves were successively taken with $I_\mathrm{max}=1$,~3,~10, and 40~mA.}
\label{fig3}
\end{figure}

Next, we show another novel memory effect due to Joule self-heating in the field-induced PS state. Figure~3(a) shows $V-I$ curves at 8~T obtained after ZFC to 4.2~K. In these measurements, the maximum current ($I_\mathrm{max}$) was increased successively from 10~mA to 40~mA. The meaning of the data becomes clearer when we plot them into $V/I$~vs.~$I$ [Fig.~3(b)]. In this data set, we successively changed the $I_\mathrm{max}$ from 1~mA to 40~mA to show this effect in a logarithmic scale. Each curve in the current-decreasing process sustains a certain value determined by $I_\mathrm{max}$. In the successive current-increasing process, $V/I$ is constant until the current exceeds the previous $I_\mathrm{max}$. In other words, the sample remembers the states at $I_\mathrm{max}$ even after reducing the current to zero. We ascribe this phenomenon to the Joule heating, not to the voltage inducing breakdown of the COI state~\cite{Asamitsu}. The right axis of Fig.~3(b) represents the effective sample temperature $T^*$ estimated by the temperature dependence of resistivity at 8~T in the warming process after ZFC [see Fig.~1(a)]. It is reasonable that the maximum power of $\sim 2$~mW at $I = 40$~mA can heat up the sample from 4.2~K to  9~K.

We can explain these nontrivial phenomena qualitatively using a phenomenological model~\cite{Voloshin}. After ZFC most parts of the crystal changes into the COI state at low temperatures. Application of a magnetic field lowers the free energy of the FMM state. In the present sample, however, not the whole sample turns into the FMM state due to the distribution of the potential barrier between the two states. Joule heating stimulates local transitions to the FMM state over the energy barrier, and shortens the lifetime $\tau$($H$,$T$), which determines the relaxation rate from the COI state. Accordingly, in the current-increasing process, a remarkable relaxation to the FMM state takes place. On the other hand, in the current-decreasing process, decrease in the sample temperature makes $\tau$ longer. When $\tau$ is longer than the measurement time, no change is observed, because the FMM regions are frozen.

\begin{figure}[!tb]
\includegraphics[width=7.5cm]{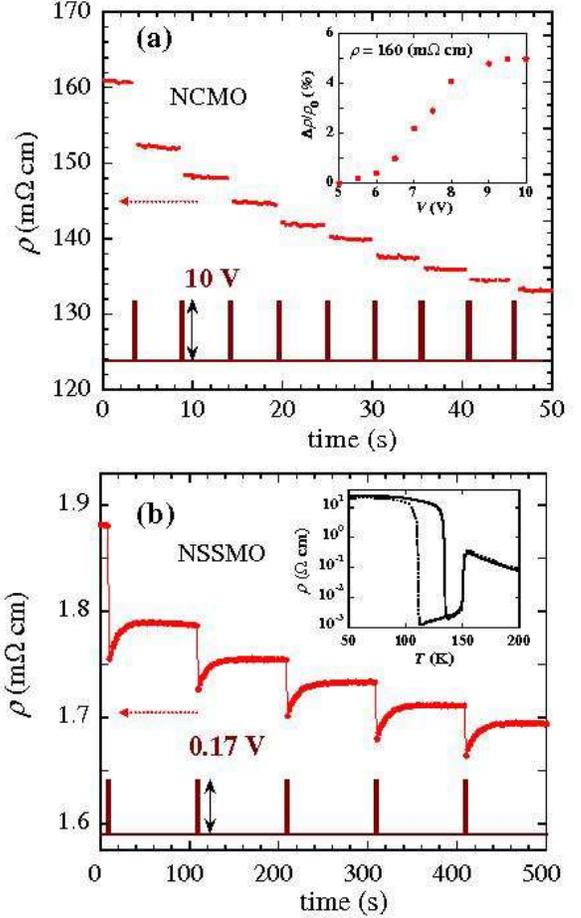}
\caption{(a) Variation of resistivity in NCMO caused by pulsed-voltages with amplitude~/~duration of 10~V~/~50~ms at 13~K and 7.5~T. The inset shows the drop of resistivity ($\Delta\rho$) from the initial value $\rho_0$ (160~m$\Omega$cm) caused by single voltage-pulse with various amplitudes. (b) Results in NSSMO by pulsed-voltages with 0.17~V~/~500~ms at 130~K and 0~T. The inset shows temperature dependence of resistivity in the NSSMO. The solid (dashed) lines indicate resistivity measured in the warming (cooling) process.}
\label{fig4}
\end{figure}

Up to now, we have discussed the way to control the fraction of one phase in the PS state using hysteretic behavior against changes in external field and temperature. In the following, we show the non-volatile resistive memory behavior caused by pulsed-voltages. Figure~4(a) shows the change of resistivity in NCMO upon applying  pulsed-voltages of 10~V with a duration of 50~ms. These measurements were carried out at 7.5~T after ZFC to 13~K. As shown in this figure, the resistivity decreases step by step by a repetition of pulses with the same amplitude and duration. This change can be ascribed to the increase in the applied power for every voltage pulse, because the power is inversely proportional to the resistance at a constant voltage. In this way, multiple values of resistance can be memorized by heat pulses, 
and they can be erased by removing magnetic field because the COI state recovers at zero field 
[see inset of Fig.~1(a)]. We can also control the step height in the resistance drop by changing the amplitude of voltage pulses as shown in the inset of Fig.~4(a).

Such a memory effect is not specific to the NCMO, but is commonly observed in other materials within the hysteretic region between FMM and COI phases. In Fig.~4(b), we show a similar memory behavior observed in NSSMO at 130~K. This material is known to show a prominent hysteresis: the COI state sets in below about 110~K upon cooling, while it melts at around 130~K upon heating [inset of Fig.~4(b)]~\cite{Tokura}. In this case, a similar step-wise change in the resistance shows up at zero field and higher temperature. The recorded low resistance state can be reset by a thermal cycle. The present two materials shown here are merely prototypes to demonstrate the potential for the memory devices. It is worth mentioning that there are several reports on charge-ordering in manganites above room temperature~\cite{GarciaMunoz,Nakajima}. The operating conditon of the memory effect can be significantly improved by optimizing the material parameters.

This multivalued memory effect reminds us of another type of resistive memory effects recently reported in some transition metal oxides~\cite{Liu,Beck}. Although these effects set in at room temperature, the origin  is still controversial~\cite{Rozenberg,Sawa}. Further studies are needed to control this effect in a reproducible manner. On the other hand, our result observed in the four-probe configuration is a bulk effect, and can be easily controlled by tuning the heat pulse. Such controllability together with drastic change in resistance enables us to use them as a multivalued memory within a single memory unit.

Although the present memory effect is driven by thermal cycles, it does not necessarily mean a slow response as a device. The present phenomena are rather analogous to the memory effects in phase-change chalcogenides~\cite{Lankhorst}. In this case, crystalline and amorphous states are selectively achieved through fast thermal cycles at a rate of $10^{11}~\mathrm{K/s}$. In manganites, we utilize the first-order transition between electronic solid and liquid that require less temperature change than the chalcogenides. 
 
In summary, we have demonstrated non-volatile multivalued memory effects in bulk crystals of \NCMO\ and \NSSMO\ in the bi-stable region of competing ferromagnetic metal and charge-ordered insulating states. The essential role of the current and the voltage is interpreted as a self-heating effect, which opens a possibility to operate the multivalued memory effect in a controllable way.

\begin{acknowledgments}
This work is supported by Grant-in-aid for Scientific Research from the Ministry of Education, Culture, Sports, Science and Technology.
\end{acknowledgments}

\end{document}